\documentclass[twocolumn]{aastex631}

\shorttitle{SED Analysis for HWO Target Stars}
\shortauthors{Stephen R. Kane et al.}

\begin{document}

\title{Spectral Energy Distribution Analysis for Habitable Worlds Observatory Target Stars}

\author[0000-0002-7084-0529]{Stephen R. Kane}
\affiliation{Department of Earth and Planetary Sciences, University of California, Riverside, CA 92521, USA}
\email{skane@ucr.edu}

\author[0000-0002-5823-4630]{Kaspar von Braun}
\affiliation{Lowell Observatory, 1400 West Mars Hill Road, Flagstaff, AZ 86001, USA}
\affiliation{Carnegie Science, Earth and Planets Laboratory, 5241 Broad Branch Rd NW, Washington, DC 20015, USA}

\author[0000-0002-3551-279X]{Tara Fetherolf}
\affiliation{Department of Earth and Planetary Sciences, University of California, Riverside, CA 92521, USA}

\author[0000-0003-0595-5132]{Natalie R. Hinkel}
\affiliation{Department of Physics \& Astronomy, Louisiana State University, 202 Nicholson Hall, Baton Rouge, LA 70803, USA}


\begin{abstract}

The Habitable Worlds Observatory (HWO) will directly image and
characterize planets potentially similar to Earth orbiting nearby
stars. Accurate stellar properties, particularly effective
temperatures ($T_\mathrm{eff}$), angular diameters ($\theta_\star$),
and bolometric fluxes ($F_\mathrm{bol}$), are essential for reliable
Habitable Zone (HZ) calculations, exoplanet yield predictions, and
coronagraph design trade studies. We present a spectral energy
distribution (SED) analysis for the 164 stars from the HWO Exoplanet
Exploration Program (ExEP) list, using $\chi^2$-minimization fits of
empirical spectral templates from the Pickles stellar spectral flux
library to broadband photometric data. The SED fits provide direct
measurements of $F_\mathrm{bol}$. We adopt spectroscopic
$T_\mathrm{eff}$ values from the PASTEL catalog for 127 of the 164
stars, with the remainder drawn from the HWO ExEP catalog, and use
these in combination with $F_\mathrm{bol}$ to calculate $\theta_\star$
via the Stefan-Boltzmann equation. As a consistency check, we compare
the Pickles template $T_\mathrm{eff}$ with the adopted spectroscopic
values and find agreement to within $\sim$3\% for 65\% of the sample,
with a systematic tendency for the templates to yield cooler
values. We identify 48 stars with discrepancies exceeding 200~K,
including close binary systems where photometric contamination
compromises the SED solution. We discuss the implications for HZ
boundary calculations, atmospheric retrieval of directly imaged
planets, and target prioritization for HWO. This SED catalog
constitutes a uniform set of empirically determined $F_\mathrm{bol}$
values and associated stellar parameters for the HWO target sample.

\end{abstract}



\section{Introduction}
\label{intro}

A fundamental goal of exoplanet science is to obtain reflected-light
spectra of Earth-sized planets in the habitable zones (HZs) of nearby
stars, enabling the search for atmospheric biosignatures
\citep{kasting1993a,kopparapu2013a,meadows2018c,schwieterman2018}. The
Habitable Worlds Observatory (HWO), recommended by the Astro2020
Decadal Survey\footnote{\url{https://doi.org/10.17226/26141}} as the
next NASA flagship astrophysics mission, is designed to detect and
spectrally characterize at least 25 potentially Earth-like planets
using an onboard coronagraph to achieve the planet-star contrast ratios
($\sim$$10^{-10}$) required for such observations. The success of HWO
depends critically on accurate characterization of its target stars:
stellar luminosity ($L_\star$), effective temperature
($T_\mathrm{eff}$), and angular diameter ($\theta_\star$) directly
determine the location and extent of the circumstellar HZ
\citep{kasting1993a,kopparapu2013a,kopparapu2014,kane2012a,kane2016c,hill2023},
the expected planet-star flux ratio, and the angular separation at
which HZ planets can be observed
\citep{kopparapu2018,vaughan2023}. Uncertainties in these stellar
properties propagate directly into exoplanet yield predictions and
observing strategy optimization
\citep{harada2024b,stark2024b,harada2025}.

Several target catalogs have been assembled for HWO. The NASA Exoplanet
Exploration Program Mission Star List \citep[HWO
ExEP;][]{mamajek2024} comprises 164 hand-selected stars considered the
best targets for driving early HWO science. The HWO Preliminary Input
Catalog \citep[HPIC;][]{tuchow2024} provides a broader list of
$\sim$13,000 nearby bright stars, while the community-developed Target
Stars and Systems catalog \citep[TSS25;][]{tuchow2025a} organizes
probable HWO targets into priority tiers. Additional studies have
addressed inner working angle constraints \citep{vaughan2023}, radial
velocity reconnaissance \citep{laliotis2023}, dynamical viability of
known planetary systems \citep{kane2024d,kane2024e}, and atmospheric
retrieval frameworks \citep{young2024c}.

A persistent challenge is the heterogeneity of stellar property
measurements in these catalogs. Spectral energy distribution (SED)
fitting is a well-established technique for determining stellar
$F_\mathrm{bol}$ and $T_\mathrm{eff}$
\citep{pickles1998,vanbelle2009a,boyajian2012a,boyajian2012b,vonbraun2014}.
By fitting broadband photometry to empirical spectral templates, SED
analysis yields a direct measurement of $F_\mathrm{bol}$, which, when
combined with $T_\mathrm{eff}$ or $\theta_\star$, provides the
respective other quantity via the Stefan-Boltzmann law. In this paper,
we present SED fits for the 164 stars in the HWO ExEP list. We describe
the data sources and methodology in Section~\ref{methods}, present the
SED catalog and compare stellar parameters in Section~\ref{results},
discuss the implications in Section~\ref{discussion}, and summarize in
Section~\ref{conclusions}.


\section{Methodology}
\label{methods}


\subsection{Data Sources}
\label{data}

The target sample was drawn from the HWO ExEP \citep{mamajek2024},
publicly hosted by the NASA Exoplanet Archive (NEA; DOI:
\dataset[10.26133/NEA39]{https://doi.org/10.26133/NEA39}). The 164
stars represent the highest-priority targets for HWO direct imaging of
Earth-sized planets in the HZ \citep{tuchow2024,tuchow2025a}. For each
star, we extracted distance ($d$), apparent $V$-band magnitude,
$T_\mathrm{eff}$, $L_\star$, and estimated $\theta_\star$ from the
ExEP catalog. It is worth noting that the ExEP $\theta_\star$ values
are not independent measurements but are estimated from the catalog
$L_\star$ and $T_\mathrm{eff}$ via the Stefan-Boltzmann relation
\citep{mamajek2024}.

For the SED fitting, we compiled broadband photometry from the
literature for each target star following the procedures of
\citet{vonbraun2014} and \citet{vonbraun2017}. Accurate SED fitting
requires photometric coverage that spans the Wien peak of the stellar
flux distribution, where the sensitivity to the spectral template match
is greatest, through the Rayleigh-Jeans tail. For this reason, we draw
photometry from a broad range of well-established catalogs spanning UV
through mid-IR wavelengths, including Str\"omgren $uvby$
\citep{1998AandAS..129..431H}, Johnson $UBV$ \citep{1987AandAS...71..413M}, Geneva
photometry, Tycho-2 \citep{hog2000a}, Hipparcos
\citep{2007AandA...474..653V}, 2MASS \citep{skrutskie2006}, and IRAS/WISE,
supplemented by additional literature photometry where available
\citep[see Table~\ref{tab:mmr} and associated photometry
references;][]{vonbraun2017}. Restricting the photometric input to only
a few infrared catalogs would sacrifice coverage around the Wien peak
and reduce the number of degrees of freedom in the $\chi^2$ fit,
yielding larger uncertainties on $F_\mathrm{bol}$ and poorer template
discrimination. All photometric data are individually vetted to exclude
measurements affected by saturation, source confusion, or catalog
artifacts, following the procedures described in \citet{vonbraun2017}.
The complete set of photometry files, documenting every measurement
used for each star, is provided as supplementary data accompanying this
paper.


\subsection{SED Fitting}
\label{fitting}

The SED fitting procedure follows the methodology of
\citet{vonbraun2017,vonbraun2014,boyajian2012a,boyajian2012b,vanbelle2009a}.
In brief, we perform a $\chi^2$-minimization fit of the literature
photometry to the library of empirical stellar spectral templates from
\citet{pickles1998}, correcting for filter profiles as described in
\citet{mann2015a}. The \citet{pickles1998} library comprises 131
flux-calibrated spectra spanning 1150--25000~\AA, covering a broad
range of spectral types and luminosity classes. The SED fit directly
yields $F_\mathrm{bol}$, defined as the integral of the best-fitting
template spectrum scaled to match the photometric data. This quantity
does not depend on assumptions about stellar evolutionary models or
bolometric corrections.

A key advantage of the \citet{pickles1998} library is that the
templates are empirical, not model-dependent. Alternative approaches
using synthetic model grids (e.g., ATLAS9, PHOENIX/BT-Settl) could in
principle provide finer sampling in $T_\mathrm{eff}$, $\log g$, and
[Fe/H], but would introduce model-dependent systematics into the SED
shape. Because $F_\mathrm{bol}$ is an integral quantity dominated by
the broadband photometric data rather than the fine structure of the
template spectrum, it is robust to the choice of template library at
the percent level \citep{vonbraun2017}. We therefore adopt the Pickles
library for all fits in this work, and note that a quantitative
comparison of $F_\mathrm{bol}$ values derived from empirical versus
model templates for this sample would be a valuable future exercise.


\subsection{Adopted Effective Temperatures}
\label{teff}

Stellar $T_\mathrm{eff}$ is related to $F_\mathrm{bol}$ and stellar
angular diameter $\theta_\star$ via the Stefan-Boltzmann relation:
\begin{equation}
  F_\mathrm{bol} = \frac{1}{4} \theta_\star^2 \sigma T_\mathrm{eff}^4,
  \label{eq:sb}
\end{equation}
where $\sigma$ is the Stefan-Boltzmann constant and $\theta_\star$ is
the limb-darkened angular diameter.

To calculate $\theta_\star$ from $F_\mathrm{bol}$ via
Equation~\ref{eq:sb}, we require an independent determination of
$T_\mathrm{eff}$ for each star. Rather than adopting the coarse
$T_\mathrm{eff}$ values associated with the discrete Pickles template
grid (131 values spanning the full spectral type range), we use
high-resolution spectroscopic determinations from the PASTEL catalog
\citep{soubiran2016a}. PASTEL is a regularly updated compilation of
published spectroscopic determinations of $T_\mathrm{eff}$, $\log g$,
and [Fe/H] from the literature, and serves as the primary source for
the spectroscopic parameters in the Hypatia Catalog
\citep{hinkel2014,hinkel2016}.

We cross-matched the 164 HWO ExEP stars against PASTEL using HD
identifiers and, for stars cataloged under Bayer or Flamsteed
designations, verified alternate identifiers via SIMBAD. For each
matched star, we computed the error-weighted mean of all PASTEL
$T_\mathrm{eff}$ determinations with reported uncertainties.
Measurements without reported uncertainties were excluded from the
weighted mean but retained as consistency checks. A total of 127 stars
(77\% of the sample) have PASTEL matches with at least one
$T_\mathrm{eff}$ determination. The median number of reported PASTEL
determinations per matched star is 17, and the median error-weighted
uncertainty on the adopted $T_\mathrm{eff}$ is 40~K. We note that the
individual PASTEL determinations are drawn from heterogeneous
literature sources employing different spectra, line lists, and
analysis methods, and are not necessarily independent; the formal
weighted-mean uncertainties may therefore underestimate the true
systematic uncertainty on the adopted $T_\mathrm{eff}$ for some stars.
We do not impose an explicit uncertainty floor on the adopted
$T_\mathrm{eff}$ values, instead reporting the formal weighted-mean
uncertainties directly. However, in light of the systematic
comparisons discussed in Section~\ref{comp}, we caution that a
realistic floor on the true $T_\mathrm{eff}$ uncertainty for
well-studied stars is of order 100~K.  For the remaining 37 stars
without PASTEL determinations, we adopt the $T_\mathrm{eff}$ values
from the HWO ExEP catalog; these are identified in Table~\ref{tab:mmr}
with a table note. The adopted $T_\mathrm{eff}$ values are combined
with the SED-derived $F_\mathrm{bol}$ to calculate $\theta_\star$
using the methods of
\citet{mann2015b,boyajian2012a,boyajian2012b,vonbraun2014}. A detailed
discussion of the error analysis framework is provided by
\citet{vonbraun2017}.


\section{Results}
\label{results}


\subsection{SED Catalog}
\label{catalog}

\begin{figure*}
    \begin{center}
        \begin{tabular}{cc}
            \includegraphics[width=8.0cm]{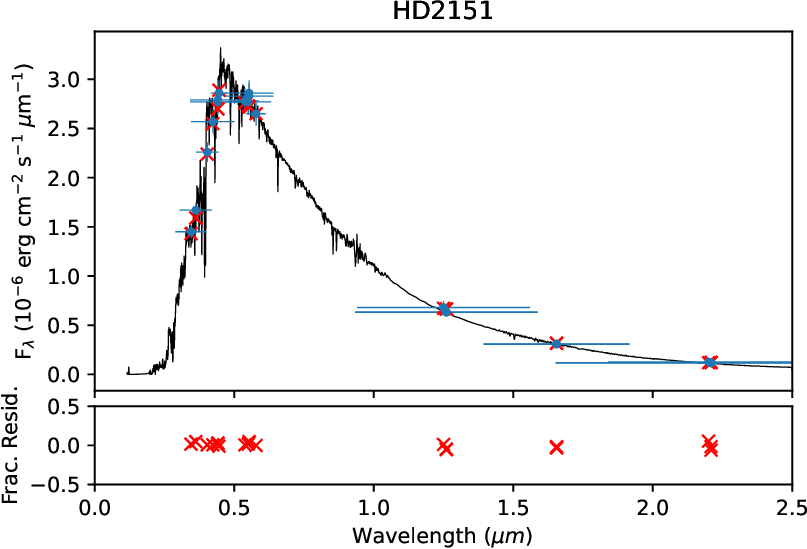} &
            \includegraphics[width=8.0cm]{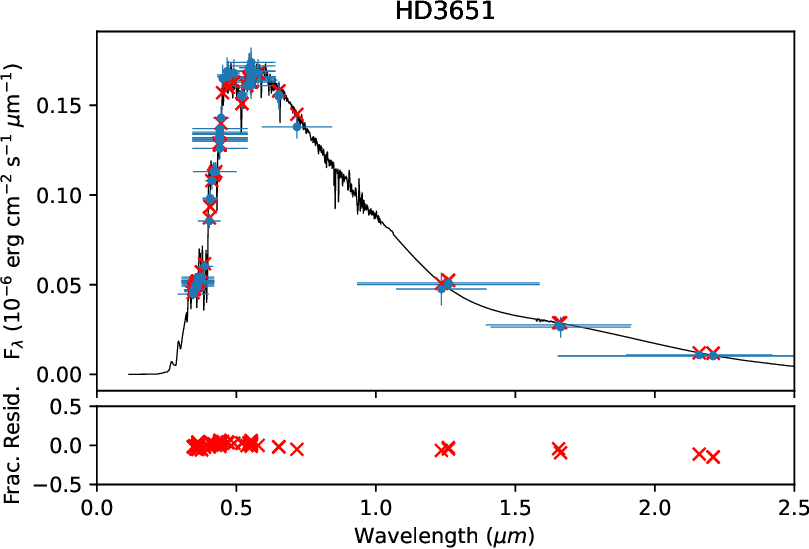} \\
            \includegraphics[width=8.0cm]{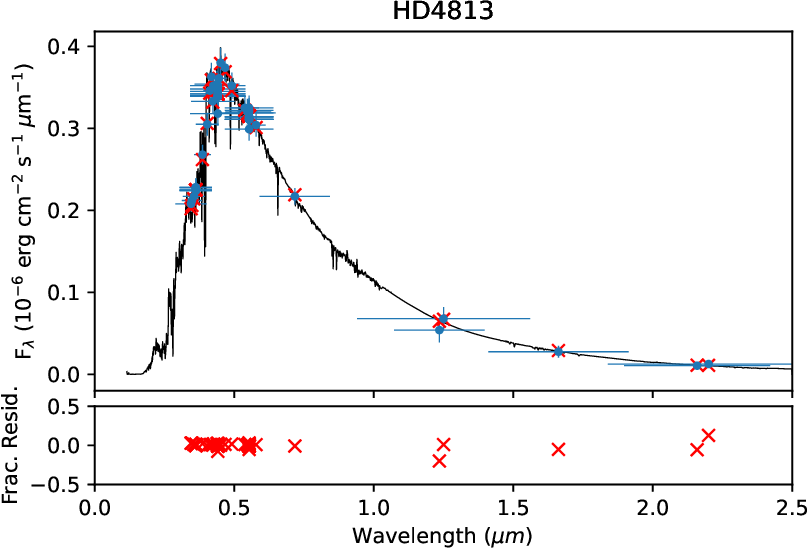} &
            \includegraphics[width=8.0cm]{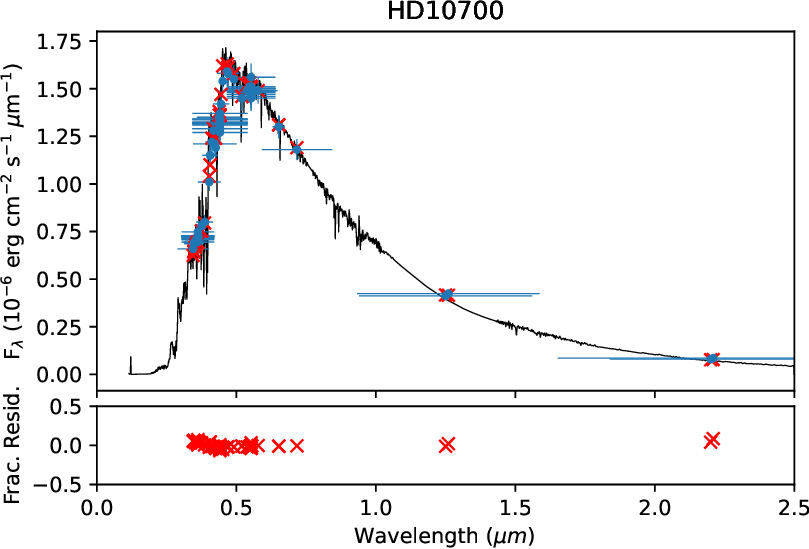}
        \end{tabular}
    \end{center}
  \caption{Representative spectral energy distribution fits for four
    HWO ExEP target stars: HD~2151 (upper left), HD~3651~A (upper
    right), HD~4813 (lower left), and HD~10700 (lower right). In each
    panel, the solid curve represents the best-fitting
    \citet{pickles1998} template spectrum, and the filled circles show
    the broadband photometry with associated uncertainties. Photometric
    error bars are shown in the $y$-direction, and filter widths in the
    $x$-direction. Fractional residuals between the template flux
    averaged over each bandpass and the literature photometry are shown
    in the lower sub-panels.}
  \label{fig:seds}
\end{figure*}

The results of the SED fitting for all 164 HWO ExEP target stars are
presented in Table~\ref{tab:mmr}. For each star, we list $d$, $V$, the
adopted $T_\mathrm{eff}$ (from PASTEL or ExEP), $L_\star$, and
$\theta_\star$ from the ExEP catalog, the Pickles template
$T_\mathrm{eff}$, the SED-integrated $F_\mathrm{bol}$, and the
calculated $\theta_\star$. Figure~\ref{fig:seds} shows representative
SED fits for four stars spanning a range of spectral types: HD~2151
($\beta$~Hydri, G2~IV), HD~3651~A (K0~V), HD~4813 (F7~IV-V), and
HD~10700 ($\tau$~Ceti, G8.5~V). In each panel, the best-fitting
\citet{pickles1998} template (solid curve) is overlaid on the broadband
photometry; photometric error bars are shown in the $y$-direction and
filter widths in the $x$-direction. The quality of the fits is
generally excellent, with reduced $\chi^2$ values near unity for the
vast majority of the sample. The complete set of individual SED fits
for all 164 stars is available as supplementary material.

The sample spans a wide range of stellar properties, from the nearby
M~dwarf HD~95735 (Lalande~21185; $T_\mathrm{eff} \approx 3460$~K,
$d = 2.55$~pc) to early F-type stars such as HD~105452~A
($T_\mathrm{eff} \approx 6600$~K, $d = 14.94$~pc). The catalog
includes well-studied benchmark systems such as
$\alpha$~Centauri~A and B, $\tau$~Ceti, $\epsilon$~Eridani, and
$\epsilon$~Indi~A \citep{difolco2007,hinkel2013a,metcalfe2013,morel2018,pathak2021,korolik2023},
as well as numerous known exoplanet host stars
\citep{akeson2013,kane2024d,christiansen2025}.

\startlongtable
\begin{deluxetable*}{l|rrrrr|r|rr}
  \tablecolumns{9}
  \tablewidth{0pc}
  \tablecaption{\label{tab:mmr} Stellar properties from the HWO ExEP catalog and derived from the SED fitting.}
  \tablehead{
    \colhead{} &
    \multicolumn{5}{c}{Adopted Stellar Properties} &
    \multicolumn{1}{c}{Template} &
    \multicolumn{2}{c}{Results} \\
    \colhead{Star} &
    \colhead{$d$} &
    \colhead{$V$} &
    \colhead{$T_\mathrm{eff}$\tablenotemark{b}} &
    \colhead{$L_\star$} &
    \colhead{$\theta_\star$} &
    \colhead{$T_\mathrm{eff}$} &
    \colhead{$F_\mathrm{bol}$} &
    \colhead{$\theta_\star$} \\
    \colhead{} &
    \colhead{(pcs)} &
    \colhead{} &
    \colhead{(K)} &
    \colhead{($L_\odot$)} &
    \colhead{(mas)} &
    \colhead{(K)} &
    \colhead{($10^{-6}$ erg cm$^{-2}$ s$^{-1}$)} &
    \colhead{(mas)}
  }
  \startdata
HD 166          & 13.77 & 6.093 & $5514\pm18$ & $0.647\pm0.010$ & 0.601 & $5322\pm186$ & $ 0.1081\pm0.0005$ & $0.592\pm0.004$ \\
HD 693          & 18.89 & 4.895 & $6196\pm50$ & $2.997\pm0.015$ & 0.741 & $6306\pm221$ & $ 0.2957\pm0.0021$ & $0.776\pm0.013$ \\
HD 739          & 21.72 & 5.241 & $6495\pm85$\tablenotemark{a} & $3.030\pm0.018$ & 0.589 & $6190\pm217$ & $ 0.2121\pm0.0011$ & $0.598\pm0.016$ \\
HD 1581         &  8.61 & 4.223 & $5959\pm48$ & $1.262\pm0.013$ & 1.149 & $5919\pm207$ & $ 0.5475\pm0.0025$ & $1.142\pm0.019$ \\
HD 2151         &  7.46 & 2.820 & $5839\pm72$ & $3.483\pm0.012$ & 2.300 & $5765\pm202$ & $ 2.0030\pm0.0218$ & $2.274\pm0.057$ \\
HD 3651 A       & 11.11 & 5.863 & $5274\pm19$ & $0.532\pm0.009$ & 0.752 & $5153\pm180$ & $ 0.1345\pm0.0006$ & $0.722\pm0.005$ \\
HD 4391         & 15.05 & 5.795 & $5917\pm55$ & $0.923\pm0.012$ & 0.571 & $5765\pm202$ & $ 0.1315\pm0.0006$ & $0.567\pm0.011$ \\
HD 4614 A       &  5.01 & 3.444 & $5907\pm12$\tablenotemark{a} & $1.252\pm0.007$ & 1.983 & $5919\pm207$ & $ 1.1290\pm0.0043$ & $1.668\pm0.007$ \\
HD 4628         &  7.44 & 5.729 & $4917\pm35$ & $0.300\pm0.011$ & 0.911 & $4976\pm174$ & $ 0.1664\pm0.0008$ & $0.924\pm0.013$ \\
HD 4813         & 15.92 & 5.176 & $6252\pm47$ & $1.781\pm0.015$ & 0.674 & $6306\pm221$ & $ 0.2238\pm0.0011$ & $0.663\pm0.010$ \\
HD 5015         & 18.80 & 4.800 & $6114\pm82$ & $3.432\pm0.008$ & 0.819 & $6026\pm211$ & $ 0.3104\pm0.0032$ & $0.817\pm0.022$ \\
HD 7570         & 15.26 & 4.966 & $6120\pm30$ & $2.004\pm0.014$ & 0.770 & $5790\pm203$ & $ 0.2761\pm0.0012$ & $0.769\pm0.008$ \\
HD 7788 A       & 23.26 & 4.912 & $6436\pm5$\tablenotemark{a} & $3.900\pm0.007$ & 0.635 & $6194\pm217$ & $ 0.2968\pm0.0014$ & $0.721\pm0.002$ \\
HD 9826 A       & 13.49 & 4.100 & $6107\pm59$ & $3.353\pm0.015$ & 1.110 & $5919\pm207$ & $ 0.6166\pm0.0030$ & $1.153\pm0.022$ \\
HD 10360        &  8.19 & 5.764 & $5026\pm49$ & $0.345\pm0.010$ & 0.881 & $5013\pm175$ & $ 0.1747\pm0.0011$ & $0.906\pm0.018$ \\
HD 10361        &  8.20 & 5.876 & $5111\pm34$\tablenotemark{a} & $0.304\pm0.011$ & 0.798 & $4976\pm174$ & $ 0.1565\pm0.0009$ & $0.830\pm0.011$ \\
HD 10476        &  7.64 & 5.241 & $5238\pm15$ & $0.454\pm0.010$ & 1.008 & $5153\pm180$ & $ 0.2508\pm0.0010$ & $1.000\pm0.006$ \\
HD 10647        & 17.35 & 5.520 & $6151\pm54$ & $1.550\pm0.014$ & 0.587 & $6026\pm211$ & $ 0.1647\pm0.0014$ & $0.588\pm0.011$ \\
HD 10700        &  3.65 & 3.496 & $5372\pm62$ & $0.506\pm0.010$ & 2.103 & $5565\pm195$ & $ 1.1430\pm0.0047$ & $2.030\pm0.047$ \\
HD 10780        & 10.04 & 5.626 & $5407\pm12$ & $0.516\pm0.008$ & 0.772 & $5347\pm187$ & $ 0.1677\pm0.0008$ & $0.767\pm0.004$ \\
HD 14412        & 12.83 & 6.336 & $5384\pm45$ & $0.446\pm0.009$ & 0.553 & $5565\pm195$ & $ 0.0846\pm0.0004$ & $0.550\pm0.009$ \\
HD 16895 A      & 11.15 & 4.100 & $6198\pm528$ & $2.246\pm0.010$ & 1.062 & $6194\pm217$ & $ 0.5818\pm0.0030$ & $1.088\pm0.188$ \\
HD 17051        & 17.36 & 5.395 & $6147\pm67$ & $1.737\pm0.014$ & 0.621 & $5790\pm203$ & $ 0.1883\pm0.0009$ & $0.629\pm0.014$ \\
HD 17206        & 14.28 & 4.465 & $6353\pm91$ & $2.617\pm0.017$ & 0.876 & $6194\pm217$ & $ 0.4267\pm0.0019$ & $0.887\pm0.025$ \\
HD 17925        & 10.36 & 6.038 & $5224\pm26$ & $0.401\pm0.010$ & 0.701 & $5153\pm180$ & $ 0.1114\pm0.0011$ & $0.670\pm0.007$ \\
HD 19373        & 10.58 & 4.050 & $5958\pm30$ & $2.230\pm0.013$ & 1.235 & $5765\pm202$ & $ 0.6608\pm0.0100$ & $1.255\pm0.016$ \\
HD 20010 A      & 14.00 & 3.800 & $6195$ & $4.250\pm0.015$ & 1.189 & $6026\pm211$ & $ 0.7269\pm0.0038$ & $1.217$ \\
HD 20630        &  9.28 & 4.850 & $5714\pm12$ & $0.857\pm0.012$ & 0.949 & $5619\pm197$ & $ 0.3244\pm0.0017$ & $0.956\pm0.005$ \\
HD 20766        & 12.04 & 5.513 & $5716\pm20$ & $0.794\pm0.012$ & 0.703 & $5765\pm202$ & $ 0.1675\pm0.0009$ & $0.686\pm0.005$ \\
HD 20794        &  6.04 & 4.258 & $5412\pm35$ & $0.654\pm0.009$ & 1.406 & $5658\pm198$ & $ 0.5509\pm0.0024$ & $1.388\pm0.018$ \\
HD 20807        & 12.04 & 5.232 & $5855\pm20$ & $1.017\pm0.012$ & 0.759 & $5765\pm202$ & $ 0.2247\pm0.0012$ & $0.758\pm0.006$ \\
HD 22001 A      & 21.78 & 4.703 & $6662\pm129$\tablenotemark{a} & $4.876\pm0.018$ & 0.708 & $6435\pm225$ & $ 0.3323\pm0.0018$ & $0.712\pm0.028$ \\
HD 22049        &  3.22 & 3.718 & $5091\pm31$ & $0.338\pm0.010$ & 2.154 & $5153\pm180$ & $ 0.9702\pm0.0038$ & $2.082\pm0.026$ \\
HD 22484        & 13.92 & 4.291 & $6033\pm13$ & $3.068\pm0.017$ & 1.084 & $5790\pm203$ & $ 0.5099\pm0.0026$ & $1.075\pm0.005$ \\
HD 23249        &  9.09 & 3.537 & $5053\pm68$ & $2.941\pm0.015$ & 2.297 & $4976\pm174$ & $ 1.2200\pm0.0061$ & $2.370\pm0.064$ \\
HD 23754        & 17.78 & 4.210 & $6685\pm80$\tablenotemark{a} & $5.069\pm0.018$ & 0.878 & $6289\pm220$ & $ 0.5217\pm0.0027$ & $0.885\pm0.021$ \\
HD 25457        & 18.71 & 5.361 & $6279\pm116$ & $2.024\pm0.016$ & 0.600 & $6306\pm221$ & $ 0.1861\pm0.0009$ & $0.599\pm0.022$ \\
HD 25998        & 21.19 & 5.522 & $6415\pm203$ & $2.305\pm0.016$ & 0.545 & $6306\pm221$ & $ 0.1628\pm0.0008$ & $0.537\pm0.034$ \\
HD 26965 A      &  5.01 & 4.415 & $5106\pm306$ & $0.410\pm0.006$ & 1.503 & $5347\pm187$ & $ 0.4989\pm0.0027$ & $1.484\pm0.179$ \\
HD 30495        & 13.24 & 5.489 & $5833\pm10$\tablenotemark{a} & $0.966\pm0.012$ & 0.676 & $5555\pm194$ & $ 0.1804\pm0.0008$ & $0.684\pm0.003$ \\
HD 30652        &  8.07 & 3.184 & $6443\pm14$\tablenotemark{a} & $2.710\pm0.008$ & 1.523 & $6277\pm220$ & $ 1.3470\pm0.0042$ & $1.532\pm0.007$ \\
HD 32147        &  8.84 & 6.202 & $4924\pm64$ & $0.290\pm0.014$ & 0.816 & $4571\pm160$ & $ 0.1131\pm0.0005$ & $0.760\pm0.020$ \\
HD 32923        & 15.92 & 4.915 & $5691\pm17$\tablenotemark{a} & $2.402\pm0.012$ & 0.931 & $5555\pm194$ & $ 0.3027\pm0.0016$ & $0.931\pm0.006$ \\
HD 33262 A      & 11.69 & 4.701 & $6151\pm41$ & $1.466\pm0.014$ & 0.846 & $6306\pm221$ & $ 0.3485\pm0.0014$ & $0.855\pm0.012$ \\
HD 33564        & 20.79 & 5.080 & $6376\pm91$ & $3.289\pm0.016$ & 0.670 & $6306\pm221$ & $ 0.2528\pm0.0013$ & $0.678\pm0.019$ \\
HD 34411        & 12.56 & 4.705 & $5889\pm14$ & $1.732\pm0.005$ & 0.947 & $5765\pm202$ & $ 0.3522\pm0.0015$ & $0.937\pm0.005$ \\
HD 35296        & 14.58 & 5.009 & $6128\pm63$ & $1.778\pm0.014$ & 0.754 & $6026\pm211$ & $ 0.2696\pm0.0016$ & $0.757\pm0.016$ \\
HD 37394        & 12.27 & 6.200 & $5287\pm91$ & $0.478\pm0.010$ & 0.639 & $5153\pm180$ & $ 0.1012\pm0.0004$ & $0.623\pm0.022$ \\
HD 38392        &  8.89 & 6.142 & $4950\pm62$\tablenotemark{a} & $0.280\pm0.009$ & 0.752 & $4804\pm168$ & $ 0.1141\pm0.0007$ & $0.755\pm0.019$ \\
HD 38393        &  8.90 & 3.596 & $6286\pm91$ & $2.344\pm0.016$ & 1.337 & $6194\pm217$ & $ 0.9522\pm0.0043$ & $1.353\pm0.039$ \\
HD 38858        & 15.21 & 5.973 & $5756\pm44$ & $0.827\pm0.010$ & 0.564 & $5555\pm194$ & $ 0.1179\pm0.0006$ & $0.568\pm0.009$ \\
HD 39091        & 18.29 & 5.666 & $6000\pm35$ & $1.533\pm0.013$ & 0.586 & $5765\pm202$ & $ 0.1519\pm0.0007$ & $0.593\pm0.007$ \\
HD 43042        & 21.78 & 5.200 & $6539\pm65$\tablenotemark{a} & $2.936\pm0.018$ & 0.570 & $6190\pm217$ & $ 0.2202\pm0.0029$ & $0.601\pm0.013$ \\
HD 43386        & 19.59 & 5.040 & $6480\pm80$\tablenotemark{a} & $2.957\pm0.017$ & 0.648 & $6289\pm220$ & $ 0.2480\pm0.0013$ & $0.650\pm0.016$ \\
HD 43834 A      & 10.21 & 5.076 & $5597\pm38$ & $0.866\pm0.011$ & 0.902 & $5472\pm192$ & $ 0.2690\pm0.0012$ & $0.907\pm0.012$ \\
HD 46588 A      & 18.20 & 5.440 & $6195\pm62$ & $1.826\pm0.015$ & 0.598 & $6306\pm221$ & $ 0.1776\pm0.0013$ & $0.602\pm0.012$ \\
HD 48682        & 16.61 & 5.252 & $6066\pm30$\tablenotemark{a} & $1.827\pm0.014$ & 0.685 & $5919\pm207$ & $ 0.2144\pm0.0010$ & $0.689\pm0.007$ \\
HD 50281        &  8.74 & 6.562 & $4719\pm52$ & $0.220\pm0.019$ & 0.731 & $4571\pm160$ & $ 0.0866\pm0.0003$ & $0.724\pm0.016$ \\
HD 50692        & 17.40 & 5.763 & $5912\pm16$ & $1.293\pm0.012$ & 0.577 & $5790\pm203$ & $ 0.1346\pm0.0007$ & $0.575\pm0.003$ \\
HD 53705        & 17.06 & 5.560 & $5808\pm30$ & $1.490\pm0.012$ & 0.661 & $5765\pm202$ & $ 0.1633\pm0.0010$ & $0.656\pm0.007$ \\
HD 55575        & 16.85 & 5.559 & $5942\pm29$ & $1.463\pm0.013$ & 0.638 & $5790\pm203$ & $ 0.1591\pm0.0008$ & $0.619\pm0.006$ \\
HD 58855        & 20.40 & 5.350 & $6349\pm25$\tablenotemark{a} & $2.445\pm0.016$ & 0.589 & $6194\pm217$ & $ 0.1915\pm0.0009$ & $0.595\pm0.005$ \\
HD 64379        & 18.33 & 5.085 & $6525\pm80$\tablenotemark{a} & $2.537\pm0.017$ & 0.632 & $6194\pm217$ & $ 0.2627\pm0.0012$ & $0.660\pm0.016$ \\
HD 65907 A      & 16.17 & 5.592 & $5976\pm35$ & $1.290\pm0.013$ & 0.605 & $5919\pm207$ & $ 0.1545\pm0.0012$ & $0.603\pm0.007$ \\
HD 69830        & 12.58 & 5.951 & $5414\pm29$ & $0.607\pm0.010$ & 0.653 & $5013\pm175$ & $ 0.1236\pm0.0008$ & $0.657\pm0.007$ \\
HD 69897        & 18.22 & 5.130 & $6259\pm103$ & $2.425\pm0.015$ & 0.674 & $6194\pm217$ & $ 0.2357\pm0.0009$ & $0.679\pm0.022$ \\
HD 72673        & 12.16 & 6.378 & $5252\pm40$ & $0.404\pm0.009$ & 0.585 & $5347\pm187$ & $ 0.0857\pm0.0005$ & $0.581\pm0.009$ \\
HD 72905        & 14.44 & 5.630 & $5893\pm14$\tablenotemark{a} & $0.973\pm0.012$ & 0.610 & $5765\pm202$ & $ 0.1525\pm0.0009$ & $0.616\pm0.003$ \\
HD 74576        & 11.19 & 6.556 & $4992\pm47$ & $0.318\pm0.010$ & 0.626 & $5013\pm175$ & $ 0.0799\pm0.0006$ & $0.621\pm0.012$ \\
HD 75732 A      & 12.59 & 5.960 & $5344\pm247$ & $0.635\pm0.011$ & 0.701 & $4976\pm174$ & $ 0.1351\pm0.0014$ & $0.705\pm0.066$ \\
HD 76151        & 16.85 & 6.008 & $5778\pm9$ & $0.971\pm0.012$ & 0.543 & $5619\pm197$ & $ 0.1105\pm0.0005$ & $0.545\pm0.002$ \\
HD 78154 A      & 20.52 & 4.809 & $6325\pm42$\tablenotemark{a} & $4.070\pm0.016$ & 0.762 & $6306\pm221$ & $ 0.3214\pm0.0016$ & $0.776\pm0.010$ \\
HD 78366        & 18.95 & 5.962 & $5996\pm50$ & $1.267\pm0.013$ & 0.513 & $5790\pm203$ & $ 0.1103\pm0.0007$ & $0.506\pm0.009$ \\
HD 82885 A      & 11.23 & 5.402 & $5523\pm22$ & $0.784\pm0.004$ & 0.802 & $5189\pm182$ & $ 0.2087\pm0.0007$ & $0.820\pm0.007$ \\
HD 84117        & 14.95 & 4.914 & $6117\pm63$ & $1.977\pm0.014$ & 0.767 & $6026\pm211$ & $ 0.2904\pm0.0016$ & $0.789\pm0.016$ \\
HD 84737        & 18.82 & 5.086 & $5893\pm13$\tablenotemark{a} & $2.795\pm0.012$ & 0.793 & $5765\pm202$ & $ 0.2474\pm0.0013$ & $0.785\pm0.004$ \\
HD 86728 A      & 14.93 & 5.378 & $5730\pm19$ & $1.378\pm0.009$ & 0.739 & $5658\pm198$ & $ 0.1989\pm0.0008$ & $0.744\pm0.005$ \\
HD 88230        &  4.87 & 6.550 & $4084\pm21$ & $0.103\pm0.009$ & 1.215 & $3991\pm140$ & $ 0.1464\pm0.0006$ & $1.257\pm0.013$ \\
HD 89449        & 21.22 & 4.792 & $6410\pm76$\tablenotemark{a} & $4.238\pm0.018$ & 0.732 & $6194\pm217$ & $ 0.3179\pm0.0014$ & $0.752\pm0.018$ \\
HD 90089 A      & 22.73 & 5.250 & $6831\pm98$ & $3.254\pm0.019$ & 0.538 & $6383\pm223$ & $ 0.2053\pm0.0007$ & $0.532\pm0.015$ \\
HD 90589        & 16.22 & 3.990 & $6905\pm80$\tablenotemark{a} & $5.162\pm0.018$ & 0.910 & $6477\pm227$ & $ 0.6320\pm0.0070$ & $0.913\pm0.022$ \\
HD 90839        & 12.95 & 4.820 & $6159\pm66$ & $1.605\pm0.011$ & 0.798 & $6306\pm221$ & $ 0.3084\pm0.0015$ & $0.802\pm0.017$ \\
HD 91324        & 22.06 & 4.897 & $6124\pm43$ & $4.403\pm0.015$ & 0.778 & $6306\pm221$ & $ 0.2953\pm0.0019$ & $0.794\pm0.011$ \\
HD 95128        & 13.89 & 5.037 & $5887\pm18$ & $1.577\pm0.012$ & 0.810 & $5765\pm202$ & $ 0.2617\pm0.0012$ & $0.809\pm0.005$ \\
HD 95735        &  2.55 & 7.421 & $3475\pm52$ & $0.020\pm0.010$ & 1.325 & $3464\pm121$ & $ 0.1031\pm0.0004$ & $1.457\pm0.044$ \\
HD 100623 A     &  9.56 & 5.956 & $5217\pm72$ & $0.370\pm0.012$ & 0.730 & $5347\pm187$ & $ 0.1245\pm0.0007$ & $0.710\pm0.020$ \\
HD 101501       &  9.58 & 5.308 & $5535\pm64$ & $0.609\pm0.006$ & 0.837 & $5472\pm192$ & $ 0.2166\pm0.0010$ & $0.832\pm0.019$ \\
HD 102365       &  9.32 & 4.893 & $5617\pm60$ & $0.844\pm0.011$ & 0.968 & $5509\pm193$ & $ 0.3045\pm0.0037$ & $0.958\pm0.021$ \\
HD 102870       & 10.93 & 3.602 & $6071\pm32$ & $3.572\pm0.006$ & 1.429 & $5790\pm203$ & $ 0.9698\pm0.0037$ & $1.464\pm0.016$ \\
HD 103095       &  9.17 & 6.427 & $4937\pm175$ & $0.212\pm0.004$ & 0.608 & $5013\pm175$ & $ 0.0842\pm0.0003$ & $0.652\pm0.046$ \\
HD 105452 A     & 14.94 & 4.025 & $6963\pm73$ & $4.453\pm0.003$ & 0.896 & $6600\pm231$ & $ 0.6150\pm0.0023$ & $0.886\pm0.019$ \\
HD 109085       & 18.24 & 4.297 & $6854\pm41$ & $4.624\pm0.022$ & 0.774 & $6300\pm221$ & $ 0.4957\pm0.0024$ & $0.821\pm0.010$ \\
HD 109358       &  8.47 & 4.260 & $5891\pm27$ & $1.151\pm0.007$ & 1.136 & $5790\pm203$ & $ 0.5238\pm0.0034$ & $1.142\pm0.011$ \\
HD 110897       & 17.56 & 5.958 & $5889\pm16$\tablenotemark{a} & $1.092\pm0.013$ & 0.532 & $6026\pm211$ & $ 0.1125\pm0.0005$ & $0.530\pm0.003$ \\
HD 114613       & 20.46 & 4.847 & $5690\pm31$ & $4.223\pm0.011$ & 0.962 & $5565\pm195$ & $ 0.3229\pm0.0012$ & $0.962\pm0.011$ \\
HD 114710       &  9.20 & 4.230 & $5970\pm43$ & $1.357\pm0.004$ & 1.092 & $5790\pm203$ & $ 0.5292\pm0.0020$ & $1.118\pm0.016$ \\
HD 114837 A     & 18.24 & 4.913 & $6231\pm87$ & $2.966\pm0.016$ & 0.752 & $6194\pm217$ & $ 0.2860\pm0.0016$ & $0.755\pm0.021$ \\
HD 115404 A     & 10.99 & 6.550 & $4843\pm107$\tablenotemark{a} & $0.302\pm0.014$ & 0.661 & $4976\pm174$ & $ 0.0741\pm0.0004$ & $0.636\pm0.028$ \\
HD 115617       &  8.53 & 4.735 & $5548\pm33$ & $0.825\pm0.006$ & 1.070 & $5565\pm195$ & $ 0.3622\pm0.0012$ & $1.071\pm0.013$ \\
HD 122064       & 10.07 & 6.488 & $4931\pm37$ & $0.291\pm0.016$ & 0.701 & $4571\pm160$ & $ 0.0950\pm0.0006$ & $0.694\pm0.011$ \\
HD 125276 A     & 17.99 & 5.872 & $6152\pm80$ & $1.208\pm0.015$ & 0.506 & $6306\pm221$ & $ 0.1206\pm0.0006$ & $0.503\pm0.013$ \\
HD 126660 A     & 14.53 & 4.052 & $6217\pm22$ & $4.031\pm0.008$ & 1.086 & $6306\pm221$ & $ 0.6341\pm0.0029$ & $1.129\pm0.008$ \\
HD 128167       & 15.76 & 4.465 & $6778\pm185$ & $3.182\pm0.018$ & 0.771 & $6477\pm227$ & $ 0.4239\pm0.0014$ & $0.776\pm0.042$ \\
HD 128620       &  1.33 & 0.002 & $5799\pm28$ & $1.521\pm0.004$ & 8.600 & $5658\pm198$ & $ 27.8400\pm0.3530$ & $8.596\pm0.099$ \\
HD 128621       &  1.33 & 1.350 & $5231\pm19$ & $0.503\pm0.006$ & 6.000 & $5153\pm180$ & $ 8.5160\pm0.1030$ & $5.842\pm0.055$ \\
HD 131156 A     &  6.75 & 4.540 & $5561\pm33$ & $0.553\pm0.011$ & 1.133 & $5013\pm175$ & $ 0.4679\pm0.0035$ & $1.212\pm0.015$ \\
HD 131156 B\tablenotemark{d}     &  6.75 & 6.979 & $4288\pm127$\tablenotemark{a} & $0.129\pm0.027$ & 0.898 & $5322\pm186$ & $ 0.4398\pm0.0017$ & $1.976\pm0.117$ \\
HD 131977       &  5.89 & 5.724 & $4681\pm19$ & $0.223\pm0.020$ & 1.157 & $4472\pm157$ & $ 0.2211\pm0.0009$ & $1.176\pm0.010$ \\
HD 134083       & 19.54 & 4.940 & $6435\pm44$\tablenotemark{a} & $3.255\pm0.017$ & 0.691 & $6289\pm220$ & $ 0.2720\pm0.0012$ & $0.690\pm0.010$ \\
HD 136352       & 14.74 & 5.655 & $5685\pm17$\tablenotemark{a} & $1.029\pm0.011$ & 0.660 & $5555\pm194$ & $ 0.1559\pm0.0011$ & $0.669\pm0.005$ \\
HD 140538 A     & 14.79 & 5.869 & $5681\pm7$ & $0.834\pm0.010$ & 0.592 & $5658\pm198$ & $ 0.1255\pm0.0006$ & $0.601\pm0.002$ \\
HD 140901 A     & 15.25 & 6.012 & $5591\pm30$ & $0.817\pm0.011$ & 0.585 & $5472\pm192$ & $ 0.1142\pm0.0007$ & $0.592\pm0.007$ \\
HD 141004       & 11.92 & 4.422 & $5884\pm7$ & $2.004\pm0.013$ & 1.058 & $5765\pm202$ & $ 0.4653\pm0.0016$ & $1.079\pm0.003$ \\
HD 142373       & 15.90 & 4.608 & $5804\pm60$ & $3.136\pm0.012$ & 1.019 & $5919\pm207$ & $ 0.3853\pm0.0017$ & $1.009\pm0.021$ \\
HD 142860       & 11.25 & 3.843 & $6237\pm39$ & $3.039\pm0.007$ & 1.215 & $6194\pm217$ & $ 0.7603\pm0.0033$ & $1.228\pm0.016$ \\
HD 143761       & 17.51 & 5.410 & $5843\pm42$ & $1.812\pm0.012$ & 0.705 & $5765\pm202$ & $ 0.1903\pm0.0009$ & $0.700\pm0.010$ \\
HD 146233       & 14.14 & 5.496 & $5809\pm16$ & $1.094\pm0.011$ & 0.685 & $5608\pm196$ & $ 0.1775\pm0.0009$ & $0.684\pm0.004$ \\
HD 147513       & 12.89 & 5.370 & $5877\pm39$ & $1.001\pm0.012$ & 0.698 & $5765\pm202$ & $ 0.1904\pm0.0010$ & $0.692\pm0.009$ \\
HD 149661       &  9.89 & 5.764 & $5293\pm8$ & $0.462\pm0.009$ & 0.769 & $5247\pm184$ & $ 0.1510\pm0.0007$ & $0.760\pm0.003$ \\
HD 155885\tablenotemark{d}       &  5.95 & 5.110 & $5171$ & $0.332\pm0.011$ & 1.134 & $5147\pm180$ & $ 0.6002\pm0.0043$ & $1.587$ \\
HD 155886\tablenotemark{d}       &  5.95 & 5.070 & $5100\pm50$ & $0.327\pm0.011$ & 1.131 & $5013\pm175$ & $ 0.5409\pm0.0035$ & $1.549\pm0.031$ \\
HD 156026       &  5.95 & 6.295 & $4476\pm24$\tablenotemark{a} & $0.158\pm0.027$ & 1.031 & $4414\pm154$ & $ 0.1321\pm0.0007$ & $0.994\pm0.011$ \\
HD 156274 A     &  8.79 & 5.472 & $5235\pm20$\tablenotemark{a} & $0.452\pm0.009$ & 0.864 & $5347\pm187$ & $ 0.1943\pm0.0007$ & $0.881\pm0.007$ \\
HD 156897 A     & 17.52 & 4.389 & $6756\pm80$\tablenotemark{a} & $4.131\pm0.018$ & 0.788 & $6410\pm224$ & $ 0.4517\pm0.0020$ & $0.807\pm0.019$ \\
HD 157214       & 14.59 & 5.385 & $5704\pm13$\tablenotemark{a} & $1.290\pm0.012$ & 0.742 & $5765\pm202$ & $ 0.1928\pm0.0015$ & $0.739\pm0.004$ \\
HD 158633       & 12.79 & 6.443 & $5293\pm33$ & $0.413\pm0.010$ & 0.554 & $5013\pm175$ & $ 0.0822\pm0.0004$ & $0.561\pm0.007$ \\
HD 160032       & 20.96 & 4.762 & $6620\pm80$\tablenotemark{a} & $4.549\pm0.017$ & 0.719 & $6383\pm223$ & $ 0.3192\pm0.0015$ & $0.706\pm0.017$ \\
HD 160691       & 15.60 & 5.124 & $5774\pm28$ & $1.899\pm0.011$ & 0.824 & $5565\pm195$ & $ 0.2526\pm0.0016$ & $0.826\pm0.008$ \\
HD 160915       & 17.65 & 4.860 & $6404\pm36$\tablenotemark{a} & $2.701\pm0.017$ & 0.703 & $6194\pm217$ & $ 0.3022\pm0.0012$ & $0.734\pm0.008$ \\
HD 165185       & 17.11 & 5.949 & $5923\pm57$ & $1.044\pm0.013$ & 0.533 & $5765\pm202$ & $ 0.1145\pm0.0006$ & $0.528\pm0.010$ \\
HD 165341 A     &  5.11 & 4.220 & $5375\pm69$ & $0.532\pm0.017$ & 1.574 & $5147\pm180$ & $ 0.7749\pm0.0037$ & $1.669\pm0.043$ \\
HD 165341 B\tablenotemark{d}     &  5.12 & 6.061 & $4393\pm149$ & $0.164\pm0.024$ & 1.224 & $5147\pm180$ & $ 0.7768\pm0.0038$ & $2.502\pm0.170$ \\
HD 165499       & 17.75 & 5.469 & $5951\pm29$\tablenotemark{a} & $1.731\pm0.013$ & 0.648 & $5790\pm203$ & $ 0.1716\pm0.0011$ & $0.641\pm0.007$ \\
HD 166620       & 11.10 & 6.377 & $5035\pm19$ & $0.363\pm0.011$ & 0.666 & $4976\pm174$ & $ 0.0921\pm0.0004$ & $0.656\pm0.005$ \\
HD 168151       & 23.16 & 4.990 & $6473\pm38$\tablenotemark{a} & $4.171\pm0.015$ & 0.652 & $6289\pm220$ & $ 0.2628\pm0.0012$ & $0.670\pm0.008$ \\
HD 182572       & 14.92 & 5.169 & $5593\pm19$\tablenotemark{a} & $1.710\pm0.010$ & 0.868 & $5347\pm187$ & $ 0.2614\pm0.0012$ & $0.895\pm0.006$ \\
HD 185144       &  5.76 & 4.672 & $5277\pm22$ & $0.434\pm0.012$ & 1.262 & $5347\pm187$ & $ 0.4069\pm0.0012$ & $1.255\pm0.011$ \\
HD 187013       & 20.99 & 5.005 & $6455\pm28$\tablenotemark{a} & $3.536\pm0.016$ & 0.666 & $6194\pm217$ & $ 0.2644\pm0.0011$ & $0.676\pm0.006$ \\
HD 187691 A     & 19.49 & 5.122 & $6124\pm30$ & $2.835\pm0.014$ & 0.711 & $5790\pm203$ & $ 0.2402\pm0.0012$ & $0.716\pm0.007$ \\
HD 189567       & 17.93 & 6.070 & $5727\pm22$ & $1.036\pm0.011$ & 0.536 & $5765\pm202$ & $ 0.1012\pm0.0005$ & $0.531\pm0.004$ \\
HD 190248       &  6.10 & 3.556 & $5605\pm49$ & $1.250\pm0.011$ & 1.827 & $5414\pm190$ & $ 0.9844\pm0.0041$ & $1.730\pm0.030$ \\
HD 190360       & 16.00 & 5.745 & $5588\pm33$ & $1.165\pm0.010$ & 0.675 & $5323\pm186$ & $ 0.1488\pm0.0009$ & $0.677\pm0.008$ \\
HD 191408 A     &  6.01 & 5.297 & $4977\pm46$ & $0.283\pm0.012$ & 1.106 & $5147\pm180$ & $ 0.2400\pm0.0012$ & $1.083\pm0.020$ \\
HD 192310       &  8.81 & 5.730 & $5107\pm41$ & $0.404\pm0.011$ & 0.864 & $4976\pm174$ & $ 0.1672\pm0.0007$ & $0.859\pm0.014$ \\
HD 193664       & 17.48 & 5.922 & $5935\pm36$ & $1.111\pm0.012$ & 0.531 & $5790\pm203$ & $ 0.1148\pm0.0005$ & $0.527\pm0.006$ \\
HD 197692       & 14.63 & 4.137 & $6638\pm80$\tablenotemark{a} & $3.530\pm0.018$ & 0.903 & $6289\pm220$ & $ 0.5700\pm0.0023$ & $0.939\pm0.023$ \\
HD 199260       & 21.10 & 5.709 & $6307\pm80$ & $1.945\pm0.016$ & 0.521 & $6306\pm221$ & $ 0.1386\pm0.0011$ & $0.513\pm0.013$ \\
HD 201091       &  3.50 & 5.211 & $4305\pm90$ & $0.142\pm0.009$ & 1.695 & $4436\pm155$ & $ 0.3905\pm0.0016$ & $1.847\pm0.077$ \\
HD 201092       &  3.50 & 6.043 & $3948\pm96$ & $0.078\pm0.015$ & 1.465 & $3943\pm138$ & $ 0.2268\pm0.0009$ & $1.674\pm0.082$ \\
HD 202560\tablenotemark{c}       &  3.97 & 6.690 & $3599\pm52$ & $0.083$ & 1.498 & $3909\pm137$ & $ 0.1569\pm0.0009$ & $1.675\pm0.049$ \\
HD 203608       &  9.26 & 4.229 & $6084\pm82$ & $1.466\pm0.014$ & 1.091 & $6306\pm221$ & $ 0.5584\pm0.0028$ & $1.106\pm0.030$ \\
HD 206860       & 18.13 & 5.942 & $5927\pm7$ & $1.136\pm0.013$ & 0.516 & $5790\pm203$ & $ 0.1116\pm0.0006$ & $0.521\pm0.002$ \\
HD 207129       & 15.56 & 5.575 & $5938\pm29$ & $1.207\pm0.013$ & 0.621 & $5790\pm203$ & $ 0.1554\pm0.0007$ & $0.612\pm0.006$ \\
HD 209100       &  3.64 & 4.674 & $4619\pm80$ & $0.222\pm0.021$ & 1.862 & $4571\pm160$ & $ 0.4861\pm0.0024$ & $1.790\pm0.062$ \\
HD 210302       & 18.46 & 4.940 & $6368\pm76$ & $2.918\pm0.016$ & 0.708 & $6306\pm221$ & $ 0.2858\pm0.0016$ & $0.722\pm0.017$ \\
HD 212330 A     & 20.34 & 5.318 & $5679\pm56$ & $2.726\pm0.012$ & 0.785 & $5619\pm197$ & $ 0.2089\pm0.0021$ & $0.776\pm0.016$ \\
HD 213845 A     & 23.02 & 5.210 & $6605\pm33$\tablenotemark{a} & $3.454\pm0.018$ & 0.573 & $6190\pm217$ & $ 0.2196\pm0.0012$ & $0.588\pm0.006$ \\
HD 215648 A     & 16.15 & 4.200 & $6193\pm92$ & $4.546\pm0.015$ & 1.066 & $6306\pm221$ & $ 0.5588\pm0.0023$ & $1.068\pm0.032$ \\
HD 216803       &  7.60 & 6.446 & $4617\pm46$ & $0.196\pm0.017$ & 0.853 & $4414\pm154$ & $ 0.1192\pm0.0009$ & $0.887\pm0.018$ \\
HD 217987       &  3.29 & 7.330 & $3676\pm46$ & $0.036\pm0.023$ & 1.330 & $3575\pm125$ & $ 0.0952\pm0.0005$ & $1.251\pm0.031$ \\
HD 219134       &  6.54 & 5.540 & $4884\pm49$ & $0.266\pm0.008$ & 1.028 & $4683\pm164$ & $ 0.2174\pm0.0006$ & $1.071\pm0.022$ \\
HD 219482       & 20.44 & 5.655 & $6282\pm40$ & $1.885\pm0.016$ & 0.528 & $6306\pm221$ & $ 0.1457\pm0.0007$ & $0.530\pm0.007$ \\
HD 219623       & 20.61 & 5.580 & $5986\pm75$ & $2.042\pm0.013$ & 0.580 & $6026\pm211$ & $ 0.1575\pm0.0008$ & $0.607\pm0.015$ \\
HD 222368       & 13.71 & 4.132 & $6241\pm54$ & $3.369\pm0.011$ & 1.079 & $6306\pm221$ & $ 0.5952\pm0.0027$ & $1.085\pm0.019$ \\
  \enddata
  \tablenotetext{a}{No PASTEL determination available; $T_\mathrm{eff}$ adopted from the HWO ExEP catalog.}
  \tablenotetext{b}{The adopted $T_\mathrm{eff}$ column lists the error-weighted mean spectroscopic value from the PASTEL catalog \citep{soubiran2016a} where available; stars for which no PASTEL determination exists are marked and use the ExEP catalog value instead. The ``Results'' $\theta_\star$ is calculated from $F_\mathrm{bol}$ and the adopted $T_\mathrm{eff}$ via Equation~\ref{eq:sb}.}
  \tablenotetext{c}{The data provided by the HWO ExEP list contains null values for the HD~202560 $T_\mathrm{eff}$ and $L_\star$ uncertainties. The template $T_\mathrm{eff}$ values are astrophysical parameters from \citet{pickles1998} associated with the best-fitting spectral template and serve as sanity checks; they do not factor into the $\theta_\star$ calculations.}
  \tablenotetext{d}{The SED fit for this component of a close binary system is likely affected by photometric contamination from the companion (see Section~\ref{comp}), and thus the derived $F_\mathrm{bol}$ and $\theta_\star$ should be treated with caution.}
  \tablecomments{Photometry sources: \citet{1965CoLPL...3...73J,
1966CoLPL...4...99J,
1967AJ.....72.1334C,
1973ARAandA..11...29M,
1974MNSSA..33...53G,
1975mcts.book.....H,
1975RMxAA...1..299J,
1976PASP...88...95C,
1978AandAS...34....1N,
1978mcts.book.....H,
1978PASP...90..429L,
1981AandAS...45....5E,
1981ApJS...45..437A,
1982AandAS...47..221R,
1982MNRAS.200..509J,
1982mcts.book.....H,
1983MNRAS.203..777A,
1984AandAS...57..357O,
1984ApJS...55..657C,
1985AandAS...59..461O,
1985AandAS...61..331O,
1985ApJS...59..197B,
1986EgUBV........0M,
1987AandAS...71..413M,
1988IRASP.C......0J,
1988mcts.book.....H,
1988iras....1.....B,
1989AandAS...81..401M,
1989ApJS...71..245K,
1991AandAS...89..415O,
1993cio..book.....G,
1993AandAS..100..591O,
1998AandAS..129..431H,
1999yCat.2225....0G,
1999MSS...C05....0H,
2001AJ....121.2148G,
2001BaltA..10..319S,
2002AandA...384..180F,
2002yCat.2237....0D,
2003AJ....126.2048G,
2003yCat.2246....0C,
2004ApJS..154..673S,
2006AandA...460..695T,
2006AJ....132..161G,
2007AandA...474..653V,
2007AJ....133.2524W,
2008ApJS..176..216A,
2009ApJ...694.1085V,
2009ApJS..180..117A,
2011ARep...55...31S,
2012AJ....143...68H,
2012yCat.1322....0Z,
2015AAS...22533616H,
2016AandA...586A..90P,
2018ApJS..238...29P,
2020yCat.1350....0G,
1969ArA.....5..303H,
1976AandA....53....1V,
1968VilOB..22....3K,
1981ApandSS..80..353S,
1991SvA....35..409K}}
\end{deluxetable*}


\subsection{Comparison of Stellar Parameters}
\label{comp}

\begin{figure*}
  \includegraphics[angle=270,width=\linewidth]{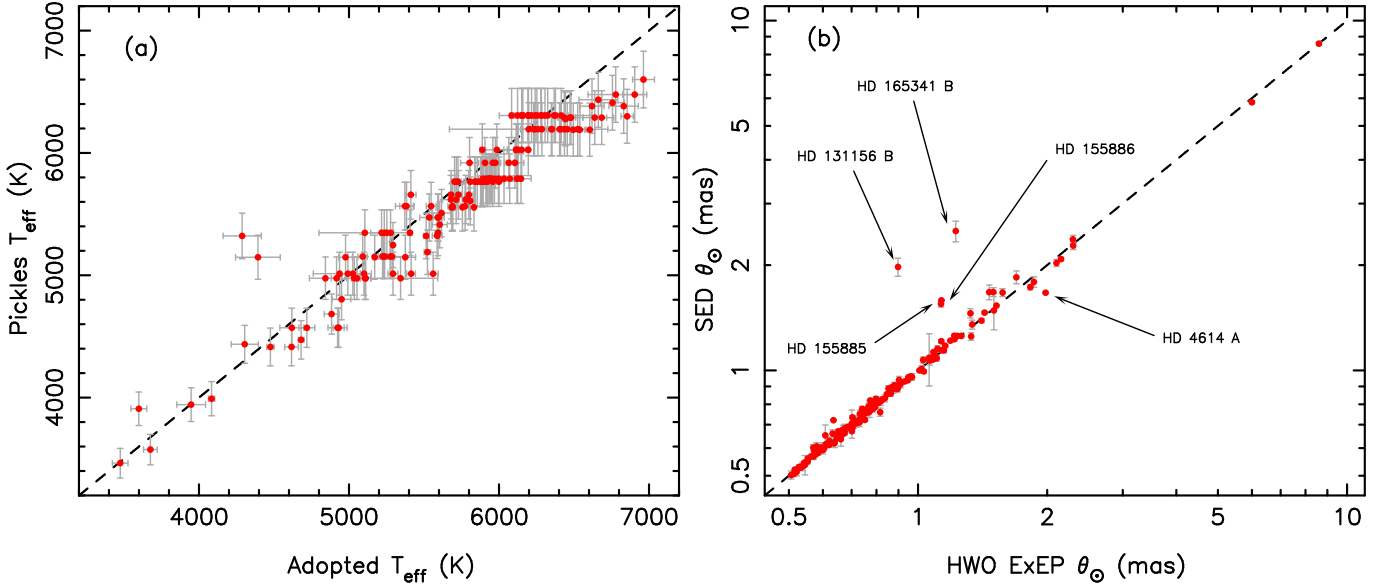}
  \caption{Panel~(a): comparison of the adopted spectroscopic
    $T_\mathrm{eff}$ (from PASTEL or HWO ExEP; see
    Section~\ref{teff}) with the best-fitting Pickles template
    $T_\mathrm{eff}$. Panel~(b): comparison of the $\theta_\star$
    values from the HWO ExEP list with those calculated from the
    SED-derived $F_\mathrm{bol}$ and the adopted $T_\mathrm{eff}$ via
    Equation~\ref{eq:sb}. Dashed lines indicate the one-to-one
    relation. Labeled systems in panel~(b) are discussed in
    Section~\ref{comp}.}
  \label{fig:comp}
\end{figure*}

A primary objective of this work is to assess the degree of agreement
between the SED-derived stellar parameters and those compiled in the
HWO ExEP list. As shown in Figure~\ref{fig:comp}(a), the majority of the
$T_\mathrm{eff}$ measurements cluster around the one-to-one line,
indicating broad consistency between the Pickles template values and
the adopted spectroscopic temperatures. The median absolute offset is
$\approx$135~K, corresponding to $\sim$2.3\% for a typical solar-type
star, and 65\% of the sample agrees to within 3\%. We note a
systematic tendency for the Pickles templates to yield cooler
$T_\mathrm{eff}$ values than the adopted spectroscopic temperatures:
76\% of the template values fall below the adopted value, with a
median signed offset of $-$112~K. This bias likely reflects the
coarse discretization of the Pickles template grid combined with
metallicity and luminosity-class effects that are not fully captured
by the 131 available templates. Despite this systematic, the level of
agreement is sufficient to confirm that the SED fits are selecting
physically appropriate templates for the majority of the sample.

The $T_\mathrm{eff}$ comparison shown in panel~(a) of
Figure~\ref{fig:comp} serves as a consistency check between two
quantities derived by entirely different methods: the adopted
spectroscopic temperatures and the best-fitting Pickles template
temperatures. It is not an independent temperature determination,
since the template temperatures are associated with a discrete grid of
131 empirical templates rather than measured for each star. We note
that the Pickles $T_\mathrm{eff}$ values are correspondingly coarser
than the spectroscopic values; the comparison is most useful for
identifying significant outliers rather than assessing precision
offsets.

A total of 48 stars (29\% of the sample) exhibit $T_\mathrm{eff}$
discrepancies $>$200~K.  Among the most prominent are the B components
of HD~131156 ($\xi$~Bootis) and HD~165341 (70~Ophiuchi). For
HD~131156~B, the adopted $T_\mathrm{eff} = 4288 \pm 127$~K, whereas
the best-fit SED template yields $T_\mathrm{eff} = 5322 \pm
186$~K. Similarly, HD~165341~B has an adopted temperature of $4393 \pm
149$~K compared to the SED value of $5147 \pm 180$~K. In both cases,
the SED-derived $F_\mathrm{bol}$ values for the B components are
nearly identical to those of their respective A components (e.g.,
$F_\mathrm{bol} = 0.4398$ vs. $0.4679$ $\times
10^{-6}$~erg~cm$^{-2}$~s$^{-1}$ for HD~131156~B and A), strongly
suggesting photometric contamination from the brighter primary
\citep{elbadry2018b}. We flag HD~131156~B and HD~165341~B as cases
where contamination is the likely source of the discrepancy.

Several additional stars exhibit $T_\mathrm{eff}$ offsets in the
150--400~K range, including HD~69830 ($\Delta T_\mathrm{eff} \approx
410$~K), HD~182572 ($\Delta T_\mathrm{eff} \approx 250$~K), and
HD~105452~A ($\Delta T_\mathrm{eff} \approx 360$~K). Discrepancies of
this magnitude are well established in the stellar characterization
literature: \citet{boyajian2012b} found systematic 200--300~K offsets
for K- and M-type dwarfs between interferometric and eclipsing binary
temperature scales, \citet{hinkel2016} documented 100--200~K scatter
between spectroscopic surveys for the same stars, and
\citet{casagrande2010} showed that the IRFM $T_\mathrm{eff}$ scale is
$\sim$100--200~K hotter than Fe excitation equilibrium values. These
comparisons indicate that $T_\mathrm{eff}$ uncertainties of order
100--200~K represent a realistic floor for many catalog values. For
stars in the present sample where the SED and spectroscopic values
diverge, follow-up characterization through interferometric angular
diameter measurements combined with SED-derived $F_\mathrm{bol}$ would
provide the definitive empirical temperature.

Panel~(b) of Figure~\ref{fig:comp} compares the $\theta_\star$
values. Because both sets of $\theta_\star$ share the adopted
$T_\mathrm{eff}$ as a common input, this comparison primarily serves as
a consistency check between our SED-derived $F_\mathrm{bol}$ and the
catalog $L_\star$. The agreement holds across the full range of angular
sizes. For the subset of stars with interferometric $\theta_\star$
measured directly with the CHARA Array
\citep{boyajian2012a,vonbraun2014}, the SED-derived $F_\mathrm{bol}$
values can be combined with those measured diameters to yield fully
empirical $T_\mathrm{eff}$ determinations independent of the adopted
spectroscopic temperature scale, although still dependent on the
SED-based $F_\mathrm{bol}$.

As labeled in panel~(b) of Figure~\ref{fig:comp}, the most significant
$\theta_\star$ outliers are associated with close binary
systems. Beyond HD~131156~B and HD~165341~B, the components of the
36~Ophiuchi system (HD~155885 and HD~155886) exhibit $\theta_\star$
discrepancies of $\sim$40\% and $\sim$37\%, respectively. This close
visual pair (separation $\sim$5$''$) presents the same photometric
contamination geometry as $\xi$~Bootis and 70~Ophiuchi. In contrast,
HD~4614~A ($\eta$~Cassiopeiae~A) presents an interesting
counterexample. The ExEP catalog lists $\theta_\star = 1.983$~mas,
while our SED analysis yields $\theta_\star = 1.668$~mas, a $\sim$16\%
offset. However, a direct interferometric measurement by
\citet{swihart2017a} gives $\theta_\mathrm{LD} = 1.623 \pm 0.004$~mas,
indicating that our SED-derived value is substantially closer to the
measured angular diameter than the ExEP catalog value. We note that
\citet{swihart2017a} employed a similar SED fitting methodology to
determine $F_\mathrm{bol}$, so the agreement is not fully independent,
but it does suggest that the ExEP $\theta_\star$ for this system may
warrant revision. These cases reinforce the diagnostic value of the
$\theta_\star$ comparison for identifying systems with compromised
photometry, even when the $T_\mathrm{eff}$ comparison alone does not
flag a discrepancy (e.g., HD~155885, whose template $T_\mathrm{eff}$
of 5147~K closely matches the adopted value of 5171~K). For HWO target
selection and yield calculations, it is important to identify and flag
such systems so that derived stellar parameters are not used
uncritically in mission planning tools.


\section{Discussion}
\label{discussion}

The boundaries of the circumstellar HZ are fundamentally determined by
$L_\star$ and $T_\mathrm{eff}$
\citep{kasting1993a,kopparapu2013a,kane2014a,kopparapu2014}. For a
solar-type star, a shift of $\sim$200~K in $T_\mathrm{eff}$ at
constant luminosity translates to a displacement of the HZ boundaries
by several percent in orbital distance, corresponding to $\sim$1--3~mas
at 10~pc \citep{kane2014a}, comparable to the anticipated inner working
angle of HWO's coronagraph \citep{vaughan2023}. For the majority of the
HWO ExEP sample, the existing catalog values are sufficiently accurate
that HZ boundary calculations are not significantly affected. However,
for systems where the $T_\mathrm{eff}$ discrepancy exceeds 200~K, the
propagated effect can be non-trivial. For example, the conservative HZ
inner boundary for HD~69830 ($d = 12.58$~pc) shifts outward by
$\sim$2\% when the SED template temperature ($T_\mathrm{eff} =
5013$~K) is adopted in place of the spectroscopic value
($T_\mathrm{eff} = 5414$~K), corresponding to $\sim$1.3~mas. Given
that HD~69830 hosts three known Neptune-mass planets
\citep{lovis2006,tanner2015}, the accuracy of the HZ delineation for
this system has concrete observational consequences.

Beyond HZ boundaries, $T_\mathrm{eff}$ uncertainties have potential
implications for the spectral characterization of directly imaged
planets. In reflected-light observations, the planetary geometric
albedo spectrum $A_g(\lambda)$ is derived by dividing out the assumed
stellar SED \citep{damiano2022a,robinson2023}. Previous studies have
shown that an error in $T_\mathrm{eff}$ introduces a
wavelength-dependent systematic: a 200~K offset for a solar-type star
alters the assumed UV flux by $\sim$20--30\% while affecting the
near-infrared by only $\sim$5\%, propagating into the retrieved albedo
spectrum as an artificial tilt that can bias the inferred strengths of
biosignature absorption features
\citep{damiano2023b,young2024c}. While we do not model these retrieval
effects directly, they emphasize the potential importance of accurate,
SED-calibrated stellar properties for HWO targets well in advance of
spectral characterization observations.

Several HWO ExEP stars are known exoplanet hosts
\citep{kane2024d,christiansen2025}, including HD~9826~A
($\upsilon$~Andromedae), HD~75732~A (55~Cancri), HD~95128
(47~Ursae~Majoris), HD~160691 ($\mu$~Arae), and HD~219134. For these
systems, the accuracy of the stellar parameters is critical for
determining whether known planets reside within the HZ
\citep{kane2012a,kane2024d,kane2024e}. The angular diameter of
HD~75732~A was previously measured interferometrically by
\citet{vonbraun2011b}, and the resulting stellar parameters were used to
characterize the system's HZ boundaries \citep{kane2011f}. The
$\sim$368~K offset between the adopted $T_\mathrm{eff}$ and the
SED template value reported here motivates a reexamination of the
catalog value in light of that interferometric measurement.


\section{Conclusions}
\label{conclusions}

We have presented a systematic SED analysis for the 164 stars in the
HWO ExEP list. Using $\chi^2$-minimization fits of empirical spectral
templates from the \citet{pickles1998} library to broadband photometric
data, we derived $F_\mathrm{bol}$ for each star. These $F_\mathrm{bol}$
values constitute the primary independent product of this work.
Combined with spectroscopic $T_\mathrm{eff}$ values adopted from the
PASTEL catalog \citep{soubiran2016a} for 127 stars and from the HWO
ExEP catalog for the remaining 37, they yield $\theta_\star$ estimates
that are broadly consistent with the catalog values. The
$T_\mathrm{eff}$ of the best-fitting Pickles templates agrees with the
adopted spectroscopic values to within $\sim$3\% for 65\% of the
sample, with a systematic tendency for the templates to yield cooler
values. A subset of 48 stars exhibit $T_\mathrm{eff}$ discrepancies
exceeding 200~K, most prominently the B
components of close binary systems where photometric contamination from
the brighter primary is the likely cause. For systems where the
temperature discrepancy is significant, the propagated shift in HZ
boundaries can exceed several percent in orbital distance. This catalog
constitutes a uniform set of empirically determined $F_\mathrm{bol}$
values for the highest-priority HWO target stars. For targets with
interferometric $\theta_\star$ from the CHARA Array or other
facilities, the $F_\mathrm{bol}$ values reported here can be combined
with the measured diameters to yield fully empirical
$T_\mathrm{eff}$ determinations independent of the adopted
spectroscopic temperature scale, although still dependent on the
SED-based $F_\mathrm{bol}$.

The principal contribution of this paper is the uniform catalog of
empirically derived $F_\mathrm{bol}$ values for the 164 HWO ExEP
target stars, determined with a single, consistent methodology. The
supplementary products accompanying this work, including the complete
photometry files for each star and the individual SED fits, are
archived in the PASP supplementary material system. Together with the
adopted spectroscopic temperatures and derived angular diameters,
these products provide a resource for HWO mission planning, target
prioritization, and the interpretation of future direct-imaging
observations.


\section*{Acknowledgments}

We would like to thank the anonymous referee whose feedback greatly
improved the manuscript, and also Gerard van Belle for very helpful
advice on effective temperature selections. This research has made use
of the NASA Exoplanet Archive, which is operated by the California
Institute of Technology, under contract with the National Aeronautics
and Space Administration under the Exoplanet Exploration Program. This
research has also made use of the Habitable Zone Gallery at
hzgallery.org, the PASTEL catalog hosted by VizieR at CDS, and the
SIMBAD database operated at CDS, Strasbourg, France. The results
reported herein benefited from collaborations and/or information
exchange within NASA's Nexus for Exoplanet System Science (NExSS)
research coordination network sponsored by NASA's Science Mission
Directorate.




\end{document}